# Two-dimensional terahertz magnetic resonance spectroscopy of collective spin waves

Jian Lu[1, †], Xian Li[1, †], Harold Y. Hwang[1], Benjamin K. Ofori-Okai[1], Takayuki Kurihara[2], Tohru Suemoto[2] and Keith A. Nelson[1, *]

[1]Department of Chemistry, Massachusetts Institute of Technology, Cambridge, MA 02139, USA.

[2]Institute for Solid State Physics, The University of Tokyo, Kashiwa, Chiba 277-8581, Japan.

†These authors contributed equally to this work.

*Corresponding author: Keith A. Nelson (kanelson@mit.edu).

**Nonlinear manipulation of spins is the basis for all advanced methods in magnetic resonance including multidimensional nuclear magnetic and electron spin resonance spectroscopies[1, 2], magnetic resonance imaging, and quantum control over individual spins[3]. The methodology is facilitated by the ease with which the regime of strong coupling can be reached between radiofrequency or microwave magnetic fields and nuclear or electron spins respectively, typified by sequences of magnetic pulses that control the magnetic moment directions[1-3]. The capabilities meet a bottleneck, however, for far-infrared magnetic resonances characteristic of correlated electron materials, molecular magnets[4], and metalloproteins[5]. Here we develop two-dimensional terahertz magnetic resonance spectroscopy and apply it to directly observe the nonlinear responses of collective spin waves (magnons). The spectra show magnon spin echoes and 2-quantum signals that reveal pairwise correlations between magnons at the Brillouin zone center. They also show resonance-enhanced second-harmonic and difference-frequency signals. Our results open the door to multidimensional magnetic resonance spectroscopy and nonlinear coherent control of terahertz-frequency spin systems in molecular complexes, biomolecules, and materials.**

Electron spin resonances (ESR) in the 0.1-10 terahertz (THz) frequency region can reveal rich information content in chemistry, biology, and materials science[1, 2, 4, 5]. In molecular complexes and metalloproteins, THz-frequency zero-field splittings (ZFS) of high-spin transition metal and rare earth ions show exquisite sensitivity to ligand geometries, providing mechanistic insight into catalytic function[5]. With strong applied magnetic fields (~ 10 tesla), resonances of unpaired electron spins in molecular complexes can be shifted from the usual microwave regime into the THz range, drastically improving the resolution of spectral splittings due to interactions with nuclear spins that reveal molecular structure[2]. In many ferromagnetic (FM) and antiferromagnetic (AFM) materials, intrinsic magnetic fields in the same range put collective spin wave transitions in the THz range. Current ESR spectroscopy remains limited at THz frequencies because the weak sources used only permitting measurements of free-induction decay (FID) signals that are linearly proportional to the excitation magnetic field strength. In some cases, including most proteins, even linear THz frequency ESR signals may not be measurable because the THz spectrum includes much stronger absorption features due to low-frequency motions of polar segments[6]. However, the fast dephasing of such motions ensures that they would not compete with nonlinear spin echo signals[7]. Two-dimensional (2D) THz ESR spectroscopy, like 2D ESR at lower frequencies, could provide extensive insight into spin interactions and dynamics in these systems. The extension of established, commercially available methodologies in multidimensional magnetic resonance spectroscopy to the THz frequency range will find a wide range of applications spanning multiple disciplines.

Collective spin excitations (magnons) in materials with spin order such as FM and AFM phases have been studied with continuous-wave and pulsed THz fields, revealing the magnon frequencies through their FID signals[8, 9] and demonstrating linear superposition in the responses to time-delayed pulse pairs[9, 10]. Nonlinear coherent control and spectroscopy could unravel complex spin interactions in AFM-to-FM order switching[11], colossal magnetoresistance[12], and multiferroicity[13], and could enable new applications in quantum computing[14], nonvolatile memory[15] and spintronics[16] at unprecedented time scales[17, 18]. THz nonlinear spectroscopy and coherent control have been demonstrated recently in various phases, almost exclusively utilizing THz electric fields[19]. So far there are very limited examples of nonlinear THz driving of spins[20] and none in which distinct nonlinear responses are separated from linear responses and from each other as is typical in 2D magnetic resonance spectroscopy. Here we explore the nonlinearity of magnons using time-delayed THz pulse pairs with magnetic field strengths in excess of 0.1 tesla. We develop 2D THz magnetic resonance spectroscopy which can be understood in terms of multiple field-spin interactions that generate the nonlinear signal fields, similar to conventional 2D magnetic resonance but in the low-order perturbative regime more typical of 2D optical spectroscopies. Magnons are resonantly excited without promoting electrons to excited states (as in most spintronics excitation) and hence the observed nonlinearities are of purely magnetic origin.

The material under study is YFeO$_3$ (YFO). The ground state has canted AFM order[8, 21] with a net magnetization along crystal *c* axis as shown in Fig. 1a. Two THz-active magnon modes, the quasi-AFM (AF) and quasi-FM (F) modes, can be constructed based on different cooperative motions of sublattice spins. Macroscopically the AF mode corresponds to oscillation of the magnetization amplitude and the F mode to precession of the magnetization orientation. In YFO at room temperature, the AF mode at frequency $f_{AF}$ = 0.527 THz and the F mode at $f_F$ = 0.299 THz can be selectively excited by THz pulses with magnetic field polarizations parallel and perpendicular respectively to the net magnetization direction. Upon THz excitation, magnons radiate FID signals **B**(*t*) at their resonance frequencies, revealing the linear-response spin dynamics in the material. We measured the FID signal fields through time-dependent electro-optic sampling (EOS) as described in the Methods section. The linear measurements yielded the data shown in Figs. 1b and 1c.

Figures 2a and 2b show the schematic experimental geometry. Two collinearly propagating single-cycle THz magnetic fields **B**$_A$ and **B**$_B$ were focused onto the sample. Rotating the sample such that the crystal *c* axis is parallel (Fig. 2a) or perpendicular (Fig. 2b) to THz magnetic fields' polarization allows selective excitation of either mode. In each geometry, incrementing the time delay $\tau$ between the two excitation pulses, at each delay we recorded the coherent time-dependent signal field **B**($\tau$,*t*) emerging from the sample by EOS. We implemented a differential chopping detection scheme[22] (see Methods and Supplementary Information 1) to isolate the nonlinear signal resulting from magnons interacting with both THz pulses. The nonlinear signal field **B**$_{NL}$ is expressed as

$$\mathbf{B}_{NL}(\tau, t) = \mathbf{B}_{AB}(\tau, t) - \mathbf{B}_A(\tau, t) - \mathbf{B}_B(t), \quad (1)$$

where **B**$_{AB}$ is the signal with both THz pulses present, and **B**$_A$ and **B**$_B$ are signals with THz pulse A and B present individually. The experimental nonlinear 2D time-domain traces **B**$_{NL}$ for each magnon mode in YFO are displayed in Figs. 2c and 2d. Oscillations along the $\tau$ axis reveal the dependence of the nonlinear signal on particular frequency components of the incident THz fields, and oscillations along the *t* axis reveal the frequency components in the nonlinear signal field. 2D Fourier transformation with respect to $\tau$ and *t* yields 2D spectrum as functions of excitation frequency $\nu$ and detection frequency $f$ as shown in Figs. 3a and 3c for each magnon mode. The difference in phase accumulation of the induced magnon coherence during time periods $\tau$ and $t$ allows the 2D spectra to be separated into nonrephasing (NR) and rephasing (R) quadrants with correspondingly positive and negative excitation frequencies (see Supplementary Information 4). Along zero excitation frequency there is a pump-probe (PP) spectral peak which results from THz-field-induced magnon population without phase accumulation.

The 2D spectrum for each magnon mode shows the complete set of third-order nonlinear signals which are observed experimentally for the first time in THz-frequency spin systems. The R (spin echo) peak and NR peak each result from a single field interaction during pulse A that creates a first-order magnon coherence and, after delay $\tau$, two field interactions during pulse B that generate a magnon population and then a third-order magnon coherence (either phase-reversed or not relative to the first-order coherence) that radiates the nonlinear signal. The 2Q peak arises from two field interactions during pulse A that create a 2-quantum coherence (2QC) accumulating phases at twice the magnon frequency and, after time $\tau$, one field interaction during pulse B that induces transitions to a third-order 1-quantum coherence that radiates the signal. The 2Q signal reveals correlations between pairs of zone-center magnons[23] (distinct from zone-boundary magnon correlations revealed in 2-magnon Raman spectra[24]). The pump-probe signal is generated by two field interactions during pulse A that create magnon population and, after delay $\tau$, one interaction with pulse B that generates a third-order coherence that radiates the signal. In addition, second-order spectral peaks due to THz second harmonic generation (SHG) and THz rectification are also identified in the AF mode 2D spectrum. The orders of the nonlinear signals are further confirmed by field dependence measurements (see Supplementary Information 3). These spectral peaks are represented concisely by the Feynman pathways shown in the Supplementary Information 4 and Fig. S7. Second-order nonlinear processes are ordinarily forbidden in a centrosymmetric crystal, but are allowed in YFO because the presence of magnetic order breaks time-reversal symmetry.

To gain further insight into the underlying physics we performed simulations based on the Landau-Lifshitz-Gilbert (LLG) equation (see Supplementary Information 4). The Hamiltonian[25] is written as

$$H = J\mathbf{S}_1 \cdot \mathbf{S}_2 + \mathbf{D} \cdot (\mathbf{S}_1 \times \mathbf{S}_2) - \sum_{i=1}^{2}(K_a S_{ia}^2 + K_c S_{ic}^2) - \gamma\big(\mathbf{B}_A(\tau,t) + \mathbf{B}_B(t)\big) \cdot \sum_{i=1}^{2} \mathbf{S}_i. \quad (2)$$

The first term describes the AFM coupling between neighboring spins $\mathbf{S}_1$ and $\mathbf{S}_2$ with a positive exchange constant $J$. The second term derives from the Dzyaloshinskii-Moria (DM) interaction with the antisymmetric exchange parameter $\mathbf{D}$ being a vector along crystal $b$ axis. The interplay between them results in the canted AFM ground state shown in Fig. 1a. The third term accounts for the orthorhombic crystalline anisotropy along crystal $a$ and $c$ axes. The last term is the Zeeman interaction between spins and THz magnetic fields with $\gamma$ being the gyromagnetic ratio. Numerically solving the LLG equation with input THz magnetic pulse pair copolarized along either crystal $c$ or $a$ axes, we obtained two vectorial solutions of the net magnetization $\mathbf{M} \propto \mathbf{S}_1 + \mathbf{S}_2$. We extracted the nonlinear temporal responses of $|\mathbf{M}|$ with THz magnetic fields polarized along crystal $c$ axis and $M_b$ with THz magnetic fields polarized along crystal $b$ axis corresponding to the experimental observables $\mathbf{B}_{NL}$ for the AF and F mode magnons respectively. 2D Fourier transformation yielded the simulated 2D spectrum of each magnon shown in Figs. 3b and 3d, exhibiting excellent agreement with the experimental results and confirming their magnetic origins. Comparing simulations with and without the DM interaction (i.e. canted AFM ground state versus AFM ground state), we found that the second-order responses of the AF mode were much stronger with the DM term. Analyzing the trajectories of $\mathbf{S}_1$ and $\mathbf{S}_2$ we note that the SHG and rectification signals are due to large-angle spin precessions (see Supplementary Information 5 and Fig. S8). The maximum excursion of $S_1$ and $S_2$ from the orientation at equilibrium is estimated to be ±0.5 degree under our THz magnetic field strengths of 0.1 tesla. Magnetization-induced SHG has been observed in the microwave frequency range (for example see Fig. 4 of reference 25) but rectification signals have not been observed before. The connection between SHG and 2Q signals, which derive from the same 2-magnon coherences, is also newly observed. The 2-magnon coherences can radiate directly at twice the magnon frequency or can be projected by a third field interaction to 1QCs that radiate at the magnon frequency. Further discussion of these signals is presented in Supplementary Information 3 and 5.

We have demonstrated 2D magnetic resonance spectroscopy of THz-frequency magnons and directly observed nonlinear spin responses. The method is generalizable to other THz-frequency magnetic resonances such as spin transitions in molecular complexes and biomolecules with high ZFS. The methodology enables study of coupled degrees of freedom in various materials utilizing both the magnetic and electric fields of THz pulses. Further enhancement of the THz magnetic field by novel THz sources[27], magnetic metamaterials[20] and resonant cavities may reveal magnetic and coupled nonlinearities beyond the perturbative limit and enable THz coherent control over magnetic domain orientation[28], with promising applications in magnon spintronics[16].

**Methods**

**Experimental setup.** The laser we used was a Ti:Sapphire amplifier system (Coherent Inc.) operating at 1 kHz repetition rate with an output power of 4 W, delivering pulses at 800 nm with 100 fs duration. 95% of the laser output was split equally into two optical paths with a controlled time delay $\tau$ between them. The optical pulses were recombined in a lithium niobate crystal to generate two single-cycle THz pulses by optical rectification utilizing the tilted-pulse-front technique[29, 30]. The two time-delayed, collinearly propagating THz pulses were collimated and focused by a pair of 90-degree off-axis parabolic mirrors, resulting in field strength in excess of 0.3 MV/cm for each pulse at the sample surface. The transmitted THz fields and FID signals were collimated and refocused into a 2 mm ZnTe electro-optic (EO) crystal by another pair of parabolic mirrors. The remaining 5% of the laser pulse energy was variably delayed by $t$ with respect to THz pulse B and overlapped with the THz fields in the EO crystal. The time-dependent magnetic field profiles $B(t)$ of the THz signals were measured through the time-dependent optical birefringence induced by their associated electric fields $E(t)$ in the EO crystal, i.e. through EOS.

**Differential chopping detection.** We used two optical choppers both at a frequency of 250 Hz to modulate the two delayed optical pulses for THz generation. As shown in the Supplementary Materials Fig. S2, in successive laser shots there was generation of both THz pulses; pulse A only; pulse B only; and neither pulse. With background noise subtraction and averaging over 100 laser shots for each data point, the sensitivity of the experiment was approximately $10^{-4}$. The choppers and laser were synced to a DAQ card (National Instruments) which measured the signal from EOS[31].

**Sample details.** The YFO sample was a 2 mm thick single crystal grown by the floating-zone method. The sample was (100) oriented with the crystal $c$-axis in the surface plane as verified by x-ray Laue diffraction.

**Acknowledgements**

This work was supported in part by Office of Naval Research Grant No. N00014-13-1-0509 and DURIP grant No. N00014-15-1-2879, National Science Foundation Grant No. CHE-1111557, and the Samsung GRO program. We thank Robert G. Griffin, Mei Hong, Thach Van Can, Prasahnt Sivarajah, Samuel W. Teitelbaum, Colby P. Steiner and Yongbao Sun for stimulating discussions.

**Author contributions**

K.A.N., H.Y.H and J.L. conceived the experimental idea. H.Y.H. and J.L. designed the experiment. J.L. collected and analyzed the data. X.L. performed numerical simulation and assisted with experiment. B.K.O.-O. wrote the LabView code for data acquisition. H.Y.H. and B.K.O.-O. contributed to the modeling and data analysis. T.K. and T.S. prepared the sample and characterized the sample orientation. K.A.N. supervised the project. All authors contributed to understanding of the underlying physics and writing the manuscript.

**Competing Financial Interests Statement**

The authors declare no competing financial interests.

**Figures and Legends**

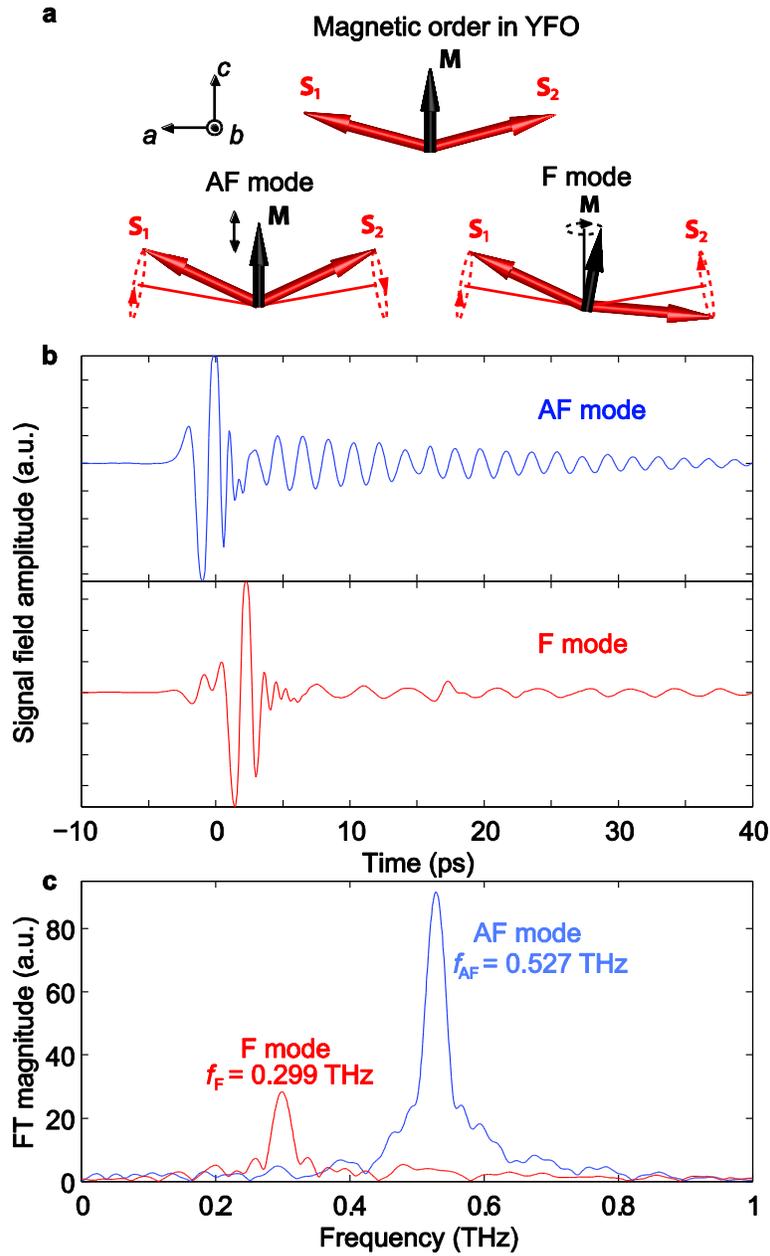

**Figure 1 | THz magnetic resonance of magnons in YFO. a,** The canted AFM order in YFO leads to a net magnetization **M** along the crystal *c*-axis. $S_1$ and $S_2$ are the $Fe^{3+}$ electron spins ordered along the crystal *a*-axis. The AF mode is the amplitude oscillation of **M** while the F mode the precession of **M**. **b,** Single THz pulses transmitted through the sample followed by FID signals from the AF (blue) and F (red) modes. **c,** Fourier transform magnitude spectra of the FID signals from the AF and F modes, showing magnon resonances at $f_{AF} = 0.527$ THz and $f_F = 0.299$ THz.

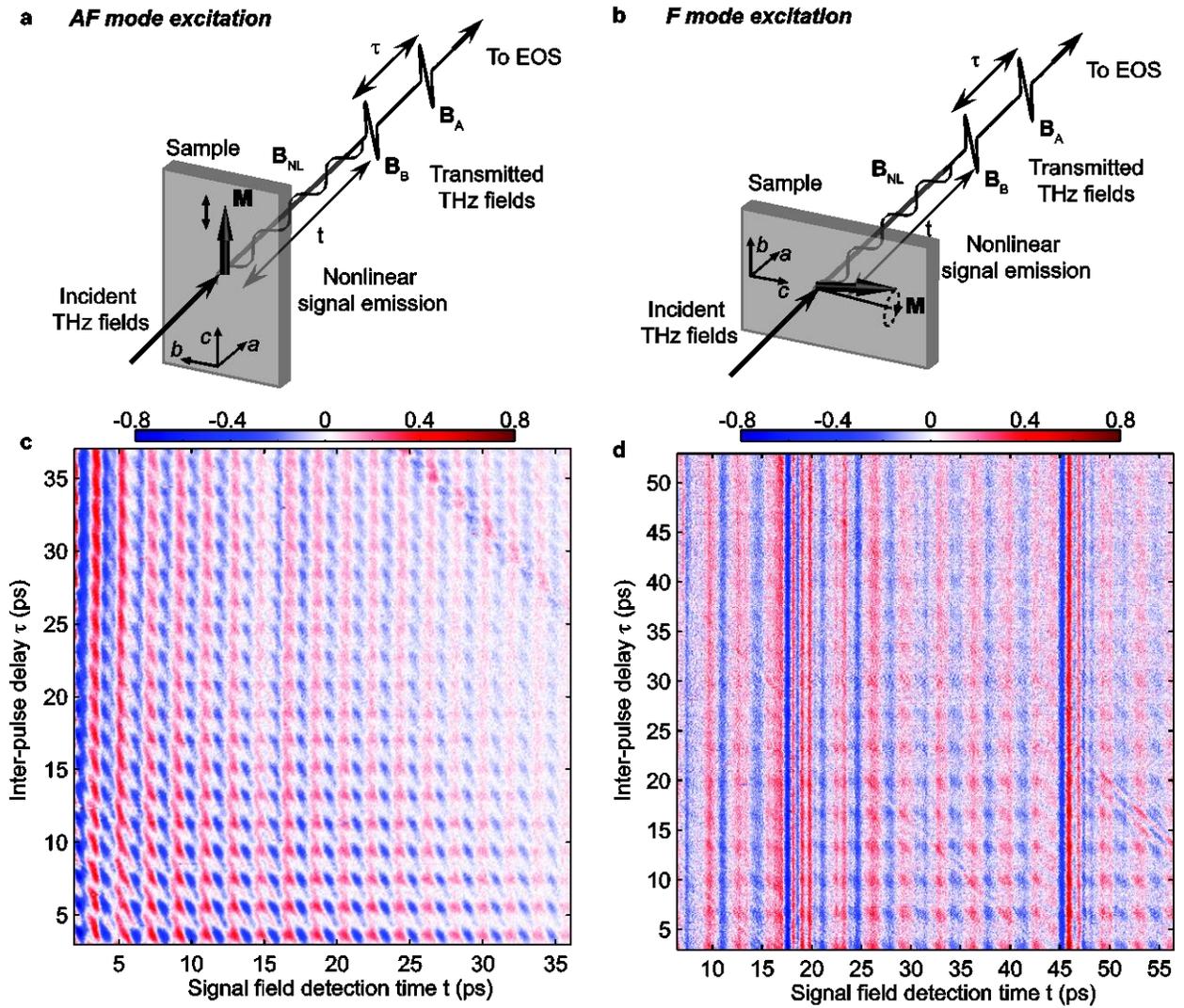

**Figure 2 | 2D THz time-domain spectroscopy. a** and **b,** Schematic illustrations of experimental geometry. Two THz pulses delayed by $\tau$ with magnetic polarization parallel or perpendicular to the sample magnetization direction **M** excite the AF (**a**) or F (**b**) mode respectively. At each inter-pulse delay $\tau$, $\mathbf{B}_{NL}(\tau,t)$ copolarized with input THz magnetic fields is measured by EOS as a function of *t*. **c** and **d,** Normalized 2D time-time plots of $\mathbf{B}_{NL}(\tau,t)$ from the AF (**c**) and F (**d**) mode respectively. Amplitudes exceeding ±0.8 are saturated in the colormap.

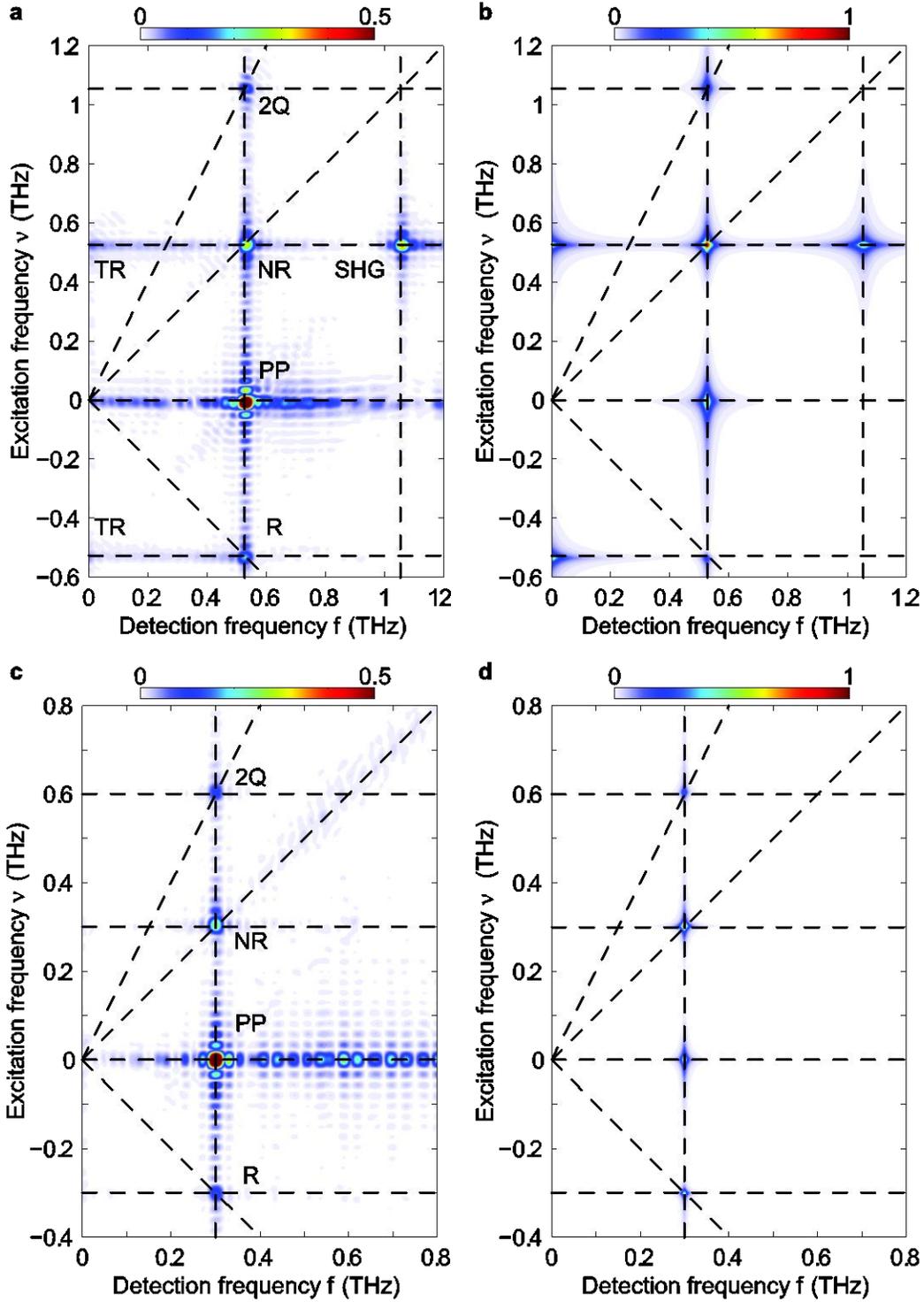

**Figure 3 | 2D magnetic resonance spectra of magnons in YFO. a** and **b,** Experimental (**a**) and simulated (**b**) AF mode 2D magnitude spectra. Third-order spectral peaks include pump-probe (PP), nonrephasing (NR) rephasing (R) and 2-quantum (2Q) peaks. Second-order peaks include second harmonic generation (SHG) and THz rectification (TR) peaks. **c** and **d,** Experimental (**c**) and simulated (**d**) F mode 2D magnitude spectra showing the full set of third-order

peaks. All spectra are normalized and plotted according to the color scale shown. Real and imaginary signals at selected frequency components are shown in Supplementary Information Fig. S4.

**Two-dimensional terahertz magnetic resonance spectroscopy of collective spin waves**

Jian Lu[1, *], Xian Li[1, *], Harold Y. Hwang[1], Benjamin K. Ofori-Okai[1], Takayuki Kurihara[2], Tohru Suemoto[2] and Keith A. Nelson[1]

[1]Department of Chemistry, Massachusetts Institute of Technology, Cambridge, MA 02139, USA.

[2]Institute for Solid State Physics, The University of Tokyo, Kashiwa, Chiba 277-8581, Japan.

*These authors contributed equally to this work.

Corresponding author: K.A.N. (kanelson@mit.edu).

**Supplementary Information**

**Table of Contents**

1. Differential chopping detection method

2. Field dependence of nonlinear spectral signals

3. Numerical simulation details

4. Double-sided Feynman diagrams

5. Origins of second-order signals

## 1. Differential chopping detection method

The time-domain waveforms of the two THz pulses used in the experiments, measured by electro-optic sampling (EOS) with a 1-mm thick GaP crystal, are plotted in Fig. S1a. The first THz pulse (A) had a peak electric field strength of 420 kV/cm characterized by EOS using a 0.1 mm GaP crystal (the thinner crystal yielded a smaller depolarized optical signal, ensuring that the signal was linearly proportional to the THz field), while the second THz pulse had a peak electric field strength of 340 kV/cm. Both THz pulses had a usable bandwidth spanning from 0.1 to 2 THz as shown in Fig. S1b. Due to the THz saturation and phase mismatch between the optical readout pulse and THz pulse in the GaP crystal, the main bandwidths were measured to be limited to 0.1-1 THz.

The pulse sequence in our differential chopping detection method is shown in Fig. S2. The laser repetition rate was 1 kHz. Two optical choppers denoted A and B were used to modulate the optical pulses that generated the corresponding THz pulses A and B at 250 Hz. On the first of four successive laser shots, the signal $\mathbf{B}_B$ resulting from THz pulse B only was measured. On the second shot, the background noise was measured with neither THz pulse A or B present. On the third shot, the signal $\mathbf{B}_A$ from THz pulse A only was measured. On the fourth shot, the signal $\mathbf{B}_{AB}$ generated by both THz pulses A and B was measured. The nonlinear signal $\mathbf{B}_{NL}(\tau,t)$ was extracted by subtracting the first three signals from the fourth at the specified inter-pulse delay $\tau$ and EOS measurement time $t$, averaging over multiple repetitions with the same time values and then scanning the values to cover the specified ranges of each.

As an example of the data taken with the above method, we show in Fig. S3 the $t$-dependent signals $\mathbf{B}_A$, $\mathbf{B}_B$, $\mathbf{B}_{AB}$ and $\mathbf{B}_{NL}$ from the AF mode in YFO at a delay $\tau = 3.7$ ps (twice the AF mode period). In the trace of $\mathbf{B}_{NL}$, a phase shift of $3\pi/2$ with respect to $\mathbf{B}_{AB}$ and asymmetric distortion are noticeable. The phase shift corresponds to the phase of the third-order nonlinear response function. The asymmetric distortion is indicative of signals from second harmonic generation, which can be resolved in the Fourier transform spectrum of the oscillatory signal in $\mathbf{B}_{NL}$, revealing peaks at $f_{AF}$ and $2f_{AF}$ as shown in Fig. S3c.

As the delay $\tau$ between the two THz pulses varies, the amplitudes and phases of the two peaks are modulated as shown in Fig. S4a, revealing their correlation to the magnon coherence induced by THz pulse A. Fourier transformation of the frequency-time plot with respect to inter-pulse delay $\tau$ yields the observed 2D spectrum of the AF mode magnons. It can be seen that the spectral peak at the AF mode fundamental frequency in Fig. S3c includes all of the R, NR, 2Q, and PP peaks in the 2D spectrum; the peak at the AF mode second harmonic frequency is the SHG peak; and the peak at zero frequency is the TR peak.

In the 2D time-domain traces in Figs 2c and 2d of the main text, observed signals due to THz pulse double reflections (from double reflections of our laser pulse in a beamsplitter delayed from each main THz pulse by ~15 ps) were not eliminated by the differential chopping detection method. But these signals did not have noticeable effects the on observed spectral responses Fourier transformed from the selected time windows. Double reflections of THz pulses in the sample (delayed from the main pulse by ~67 ps) reentered the sample and provided additional field-magnon interactions to generate stimulated nonlinear signals which were not eliminated but were excluded in the selected time windows. These stimulated nonlinear signals are beyond the scope of current study and will be explored systematically in future experiments involving three time-delayed THz pulses.

## 2. Field dependence of nonlinear spectral signals

We characterized the field dependence of the spectral amplitudes at the AF mode fundamental and second harmonic frequencies. We first measured the nonlinear time-domain traces of $\mathbf{B}_{NL}$ at a delay of $\tau = 3.7$ ps varying both THz magnetic field strengths to the same extent by attenuating the input pump laser intensity. The data are shown in Fig. S4a. The nonlinear signal excluding that at during the arrival of THz pulse B was Fourier transformed to yield the frequency-domain responses shown in Fig. S5b, where two spectral peaks at the fundamental and second harmonic frequencies of the AF mode are present. Due to the low signal-to-noise ratio around zero frequency, the THz rectification peak is not clearly discerned. We then integrated the spectrum near the peak amplitude of the spectral peaks at fundamental and second harmonic magnon frequencies and plotted them as a function of the input THz magnetic field strength as shown in Fig. S6. The integrated spectral amplitude at the fundamental frequency has a cubic dependence on the input THz magnetic field strength, while that at the second harmonic frequency shows a quadratic field dependence. The scaling behaviors further confirm that the spectral peaks observed in the 2D spectra are of third- and second-order respectively.

## 3. Numerical simulation details

A single-site, two-sublattice Hamiltonian[S1] was adapted to simulate the Brillouin-zone-center AFM resonances with a canting between two spins. Compared to a four-sublattice theory[S2], the model presents an intuitive physical picture yet is able to reasonably capture the experimental results. From the model Hamiltonian, a time-dependent effective magnetic field can be calculated by $\mathbf{B}_i^{\text{eff}} = -\frac{1}{\gamma}\frac{\partial H}{\partial \mathbf{S}_i}$ for each sublattice $i = 1, 2$ and enters the coupled LLG equation as[S3]

$$\frac{d\mathbf{S}_i}{dt} = \frac{\gamma}{1+\alpha^2}\left[\mathbf{S}_i \times \mathbf{B}_i^{\text{eff}} + \frac{\alpha}{|\mathbf{S}_i|}\mathbf{S}_i \times \left(\mathbf{S}_i \times \mathbf{B}_i^{\text{eff}}\right)\right] \quad (S1).$$

The gyromagnetic ratio $\gamma$ given by $\gamma = g\mu_B/h$, where $g = 2$ is the Landé g-factor, $\mu_B$ the Bohr magneton, and $h$ the Planck constant, is negative in our convention[S4]. The Gilbert damping parameter is chosen as $\alpha = 1.3 \times 10^{-3}$ and $5.5 \times 10^{-4}$ respectively for the AF and F modes to best match the decay time of the FID signals from the experimental result of each magnon mode. Numerically solving equation (S1) with realistic material parameters[S5, S6] as a function of time variables $\tau$ and $t$ yields the dynamics of the sublattice spins and hence of the net magnetization. The values of the parameters used in the Hamiltonian and the LLG equation for simulations are listed in Table I.

## 4. Double-sided Feynman diagrams

Double-sided Feynman diagrams representing the typical excitation pathways leading to the emission of the nonlinear signals observed in the 2D spectra are shown in Fig. S7. The time evolution is upwards. The states of the system in the diagrams are corresponding density matrix elements. Each arrow on the left side indicates a ket interaction while that on the right side indicates a bra interaction. Final signal emission is represented by a dashed arrow conventionally pointing outwards on the ket side.

We elaborate the double-sided Feynman diagram in Fig. S7a (i) as an example. Starting from the ground state $|0\rangle\langle 0|$, pulse A provides one ket interaction and induces a first-order 1-quantum coherence (1QC) $|1\rangle\langle 0|$ which evolves for time $\tau$ in the phase evolution as $\exp(i\omega_{10}\tau)$ at the 1QC frequency $\omega_{10}$ (magnon frequency). Pulse B

provides two interactions successively. The first is a bra interaction which induces a second-order population $|1\rangle\langle 1|$ and the second is a ket interaction which induces a third-order 1QC $|1\rangle\langle 0|$ evolving during the detection time period $t$ as $\exp(i\omega_{10}t)$. The third-order 1QC then emits the nonlinear signal with a total phase of $\exp(i\omega_{10}\tau + i\omega_{10}t)$ during $t$. Due to the same sign of the phase evolution of the first-order and third-order 1QCs this signal emission is termed non-rephasing (NR) signal. The difference between NR and rephasing (R) pathways is seen by comparison between Fig. S7a (i) and (ii) in which the third-order 1QC $|0\rangle\langle 1|$, with a phase of $\exp(i\omega_{01}t)$ (where $\omega_{01} = -\omega_{10}$) evolving during time $t$, is rephased with respect to the first-order 1QC $|1\rangle\langle 0|$. This rephased 1QC with a total phase of $\exp(i\omega_{01}\tau + i\omega_{10}t)$ hence reaches maximum amplitude at $t = \tau$ (neglecting relaxation processes) and emits the spin echo signal observed in the time domain. 2D Fourier transformation of the time-domain signal with respect to $\tau$ and $t$ hence can differentiate the NR and R signals by the different phase accumulation involved in these two pathways. In the 2Q pathway, pulse A provides two ket interactions successively to induce a 2QC which evolves with a phase factor $\exp(i\omega_{20}\tau) = \exp(i2\omega_{10}\tau)$. Pulse B provides one ket interaction to induce a third-order 1QC with a phase factor $\exp(i2\omega_{10}\tau + i\omega_{10}t)$. In the 2D spectra the 2Q peak is located at $\nu = 2\omega_{10}/2\pi$ and $f = \omega_{10}/2\pi$.

The subscripts of the time variables in the diagrams denote the number of preceding field-matter interactions. In the case where one pulse interacts more than once with the sample, the time variable after the last field-matter interaction corresponds to the variable time delay in the experiment. For example, in Fig. S7a (i) pulse B interacts twice with the system and hence $t_2 = 0$ and $t_3$ is the detection time $t$. Further elaboration of the double-sided Feynman diagrams can be found in reference S7.

## 5. Origins of second-order signals

It is instructive to consider the origins of the second-order signals. As shown in Fig. 1a, the AF mode involves counter-rotating magnetization components $S_1$ and $S_2$. During each cycle, both vectors reach their maxima and minima along the **M** direction (*c* crystallographic axis) at the same times, but their components along the *b* crystallographic axis always have opposite signs. The magnitude of each vector $S_i$ does not change as it precesses about its initial location. Its maximum and minimum values along *c* therefore lie equidistant from its initial location along an arc whose slope is steeper below the initial location than above it. See Fig. S8 (and for example Fig. 4 in reference S8). It therefore increases its value along *c* at its maximum by less than it decreases its value along *c* at its minimum, relative to its static value. The time-dependence of the *c*-component is not sinusoidal. It takes the form of a distorted sinusoid that includes a second harmonic component and a net DC component in the −*c* direction in Fig. S8. These radiate the second harmonic and rectification signals respectively.

Note that the second harmonic signal is derived from a 2QC, as shown in the Feynman diagram in Fig. S7b (v). A third incident field can induce a transition down to a 1-quantum coherence that radiates a third-order signal. That is the 2Q signal depicted in Fig. S7a (iii). The two diagrams show the same result after the first two field interactions. They can be viewed as resulting from interactions between magnon pairs at the Brillouin zone center. As the magnon amplitude becomes infinitesimally small, the positive and negative amplitudes along *c* become equal. At finite amplitudes, the moments of individual magnons interact, giving rise to the observed nonlinearities.

**Figures and Legends**

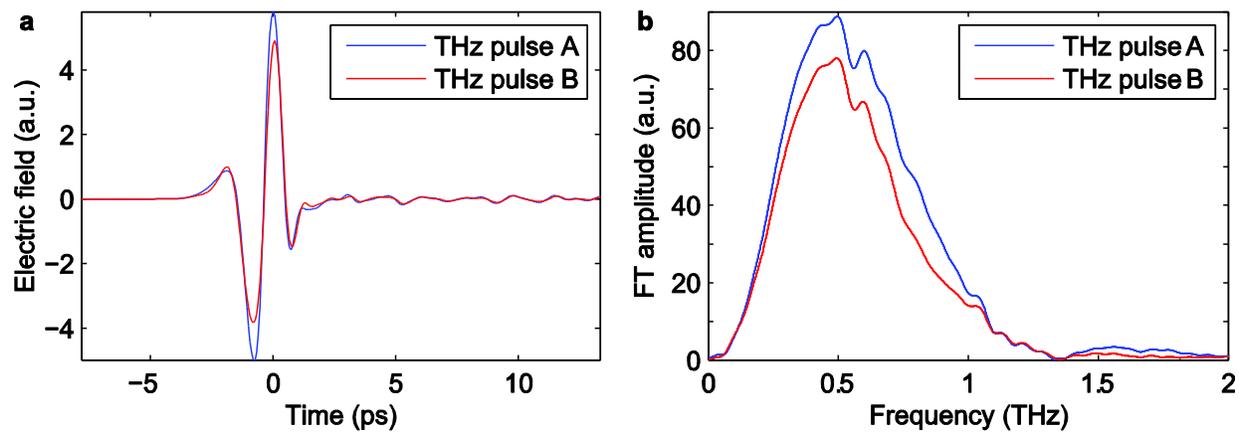

**Figure S1 | THz pulses used in experiment. a,** Time-domain waveforms of THz pulses A and B used in the experiments, measured in a 1 mm GaP EOS crystal. Due to saturation and phase mismatch in the thicker detection crystal, the peak fields in the traces are not proportional to the values characterized in a 100 μm GaP. **b,** THz spectra from a numerical Fourier transform of the waveforms in **a**.

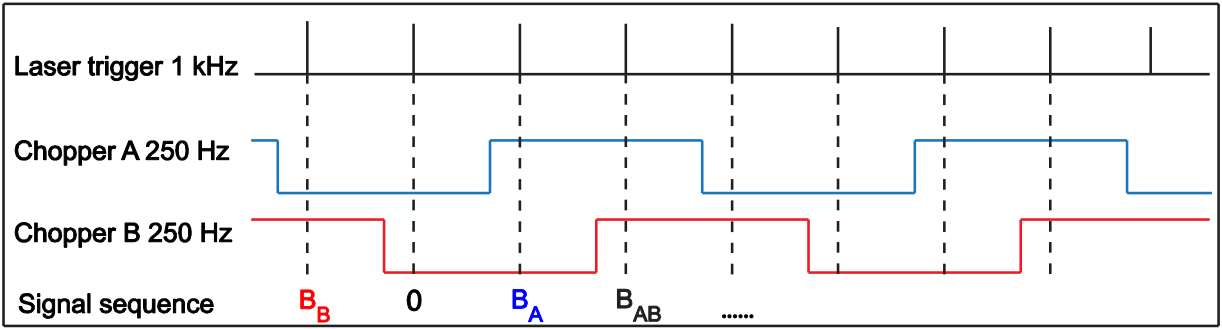

**Figure S2 | Differential chopping detection pulse sequence.** The two optical choppers at 250 Hz allow the displayed signal sequence to be generated.

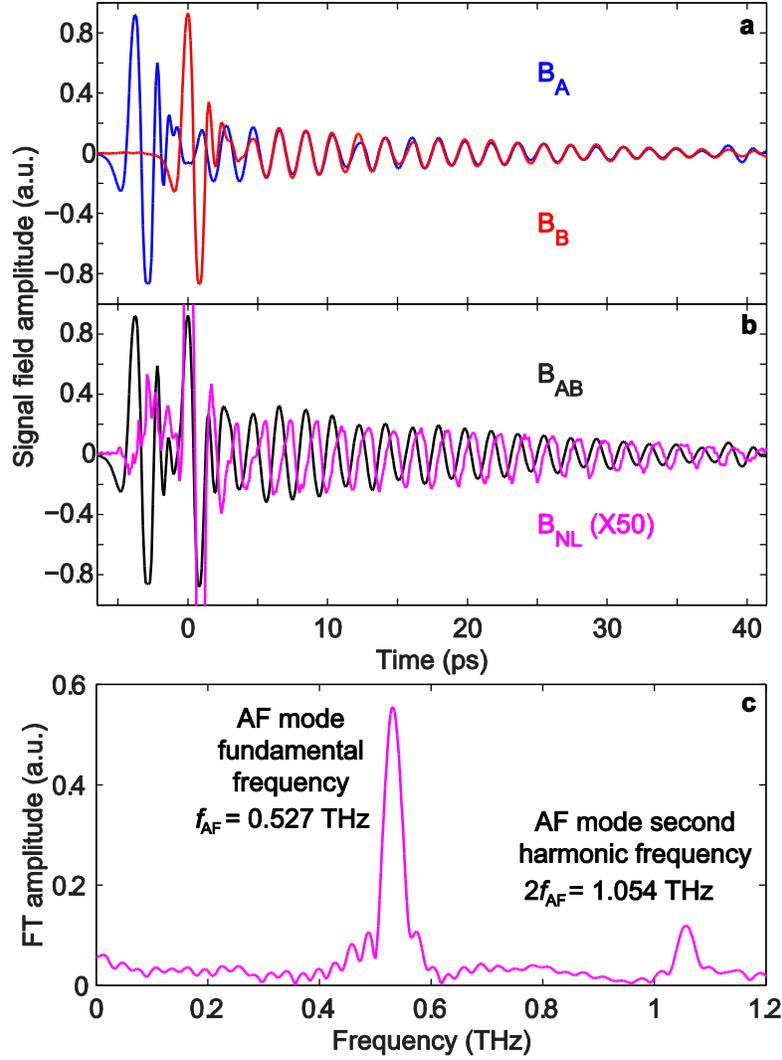

**Figure S3 | Nonlinear signal extraction from signals measured by the differential chopping detection method. a,** AF mode magnon signals induced by THz pulse A ($\mathbf{B}_A$, blue) and THz pulse B ($\mathbf{B}_B$, red) individually. **b,** Magnon responses ($\mathbf{B}_{AB}$, black) with the presence of both THz pulse A and B at a delay $\tau$ of 3.7 ps (twice the period of the AF mode) and nonlinear response ($\mathbf{B}_{NL}$ magnified 50x, magenta) extracted by $\mathbf{B}_{NL} = \mathbf{B}_{AB} - \mathbf{B}_A - \mathbf{B}_B$. $\mathbf{B}_{AB}$ is plotted for comparison. The spike during the arrival of THz pulse B is due to nonlinear absorption induced by THz pulse A (PP signal), which is excluded in all Fourier transformation in this work. **c,** Fourier transformation of the oscillatory part in $\mathbf{B}_{NL}$ in **b** reveals two spectral peaks at the fundamental and second harmonic magnon frequencies, which correspond respectively to the third- and second-order spectral peaks in the 2D spectrum.

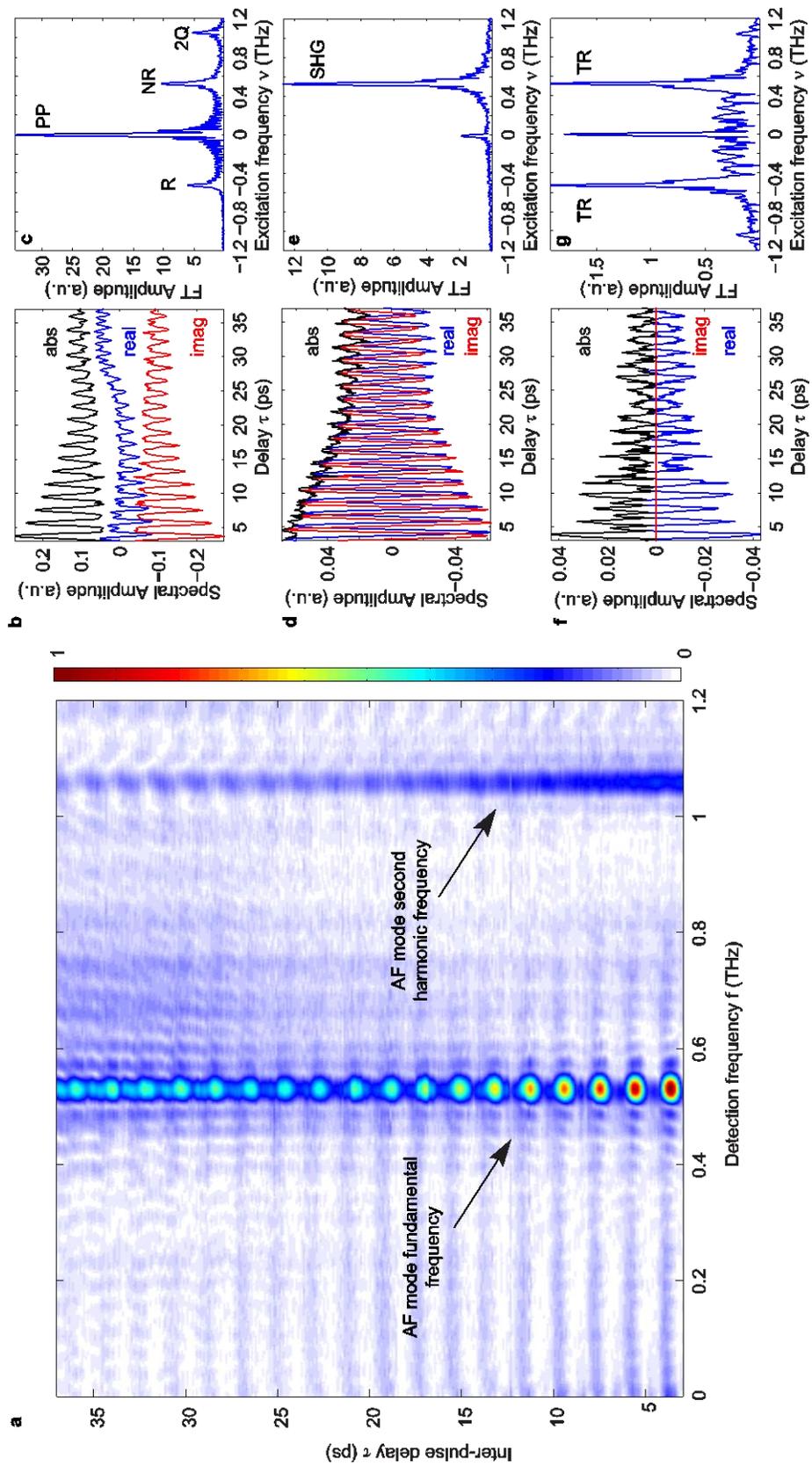

**Figure S4 | 2D time-frequency plot of AF magnon mode nonlinear response. a,** 2D time-frequency plot $\mathbf{B}_{\mathrm{NL}}(\tau, f)$ generated by a numerical Fourier transformation along $t$ of the 2D time-time plot from the AF magnon mode shown in Fig. 2a. The AF mode fundamental and second harmonic peaks are modulated as the delay $\tau$ increases. **b,** Spectral modulation as a function of $\tau$ at the AF mode frequency $f = f_{\mathrm{AF}}$. The real part (blue), imaginary part (red) and absolute value (black) of the complex spectral peak are plotted. Oscillations are clearly discernible in the time traces, showing a modulation of the spectral amplitude and phase as a function of $\tau$. **c,** Fourier transformation of the complex spectral modulation along $\tau$ in **b** yields the spectrum as a function of $\nu$, which is equivalent to taking a spectral slice along $f = f_{\mathrm{AF}}$ in the 2D spectrum in Fig. 3a. Four peaks are observed in the spectrum, corresponding to the rephasing (R), nonrephasing (NR), 2-quantum coherence (2Q), and pump-probe (PP) peaks. **d,** Spectral modulation as a function of $\tau$ along $f = f_{\mathrm{AF}}$. The real part (blue), imaginary part (red) and absolute value (black) of the complex spectral modulation are plotted as functions of $\tau$. **e,** Fourier transformation of the complex spectral peak modulation along $\tau$ in **d** yields the spectrum as a function of $\nu$, which is equivalent to taking a spectral slice along $f = 2f_{\mathrm{AF}}$, the magnon second harmonic frequency, in the 2D spectrum in Fig. 3a. **f,** Spectral modulation as a function of $\tau$ at $f = 0$ THz. The real part (blue), imaginary part (red) and absolute value (black) of the complex spectral peak are plotted as functions of $\tau$. **g,** Fourier transformation of the spectral peak modulation along $\tau$ yields the spectrum as a function of $\nu$, which is equivalent to taking a spectral slice along $f = 0$ in the 2D spectrum in Fig. 3a of the main text.

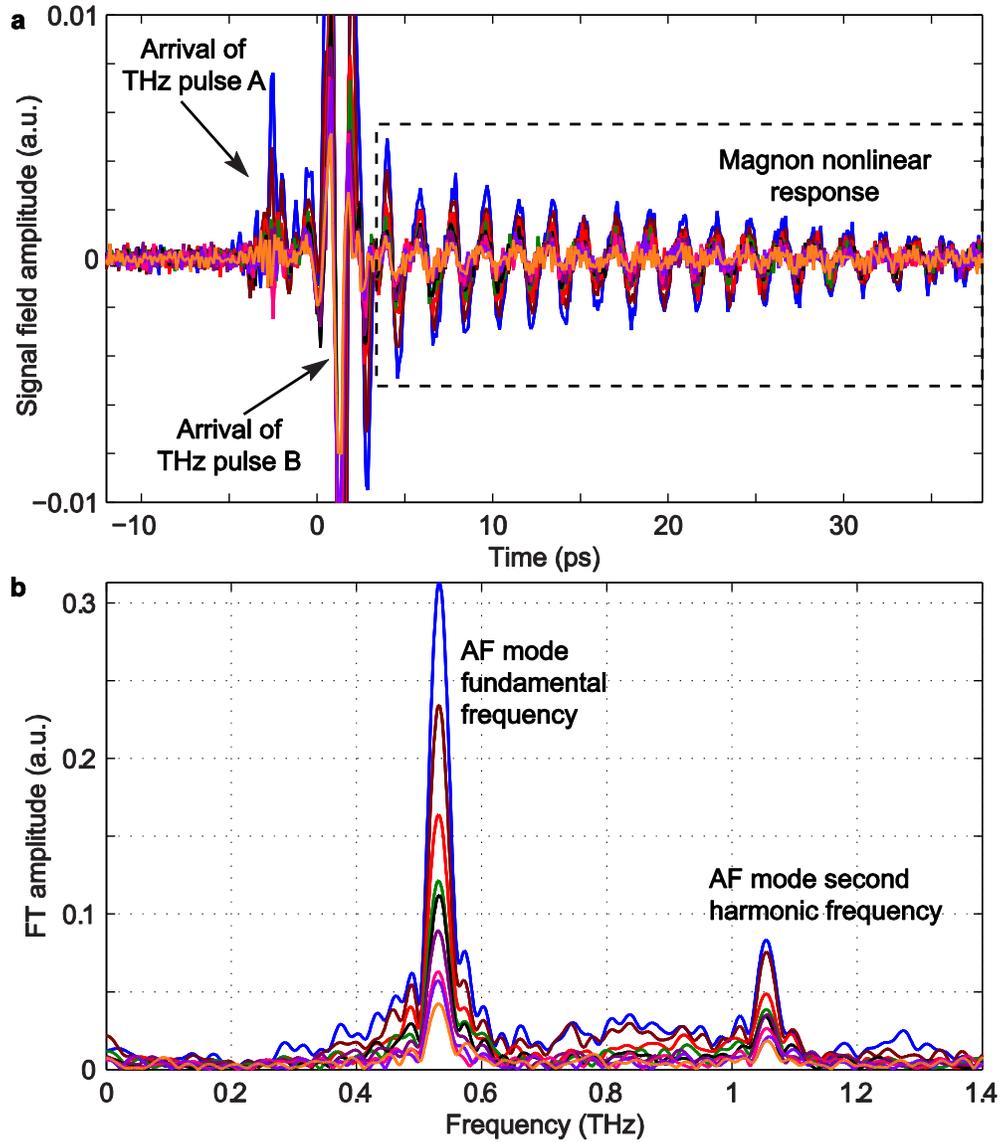

**Figure S5 | Field dependence of the nonlinear time-domain traces. a,** Time-domain traces of $\mathbf{B}_{NL}$ at varying excitation THz magnetic field strengths. The oscillatory signals in the dashed box are the nonlinear magnon responses. **b,** Fourier transformation magnitude spectra of the oscillatory signals in **a**. Two spectral peaks at the fundamental and second harmonic frequencies of the AF mode are observed. There is also a weak peak at zero frequency.

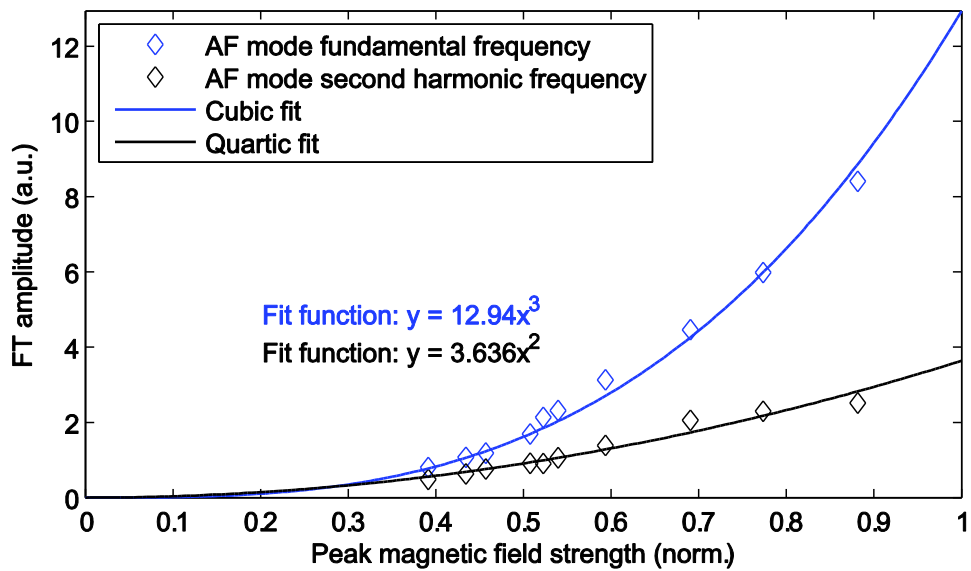

**Figure S6 | Dependence of the spectral peaks on input THz magnetic field strength.** The peaks at the AF mode fundamental and second harmonic frequencies are fitted with one fit parameter to cubic and quadratic power dependences respectively. The scaling relations with respect to the input peak THz magnetic field strength confirm the orders of the nonlinear signals observed in our experiments.

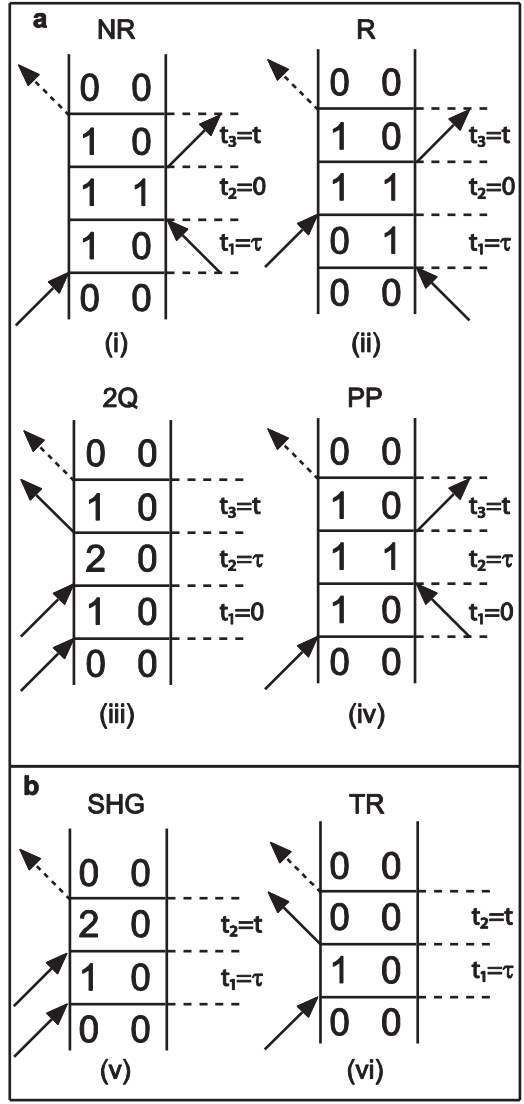

**Figure S7 | Double-sided Feynman diagrams. a** and **b,** Double-sided Feynman diagrams show typical excitation pathways leading to the coherent emission of third-order (**a**, (i)-(iv)) and second-order (**b**, (v)-(vi)) nonlinear signals. Each solid arrow denotes a field interaction that induces a transition between magnon populations or coherences denoted by diagonal ($|0\rangle\langle 0|$, $|1\rangle\langle 1|$) or off-diagonal ($|1\rangle\langle 0|$, $|2\rangle\langle 0|$) density matrix elements respectively. The dashed arrows represent the measured signal fields. The time interval subscripts denote the number of preceding field interactions.

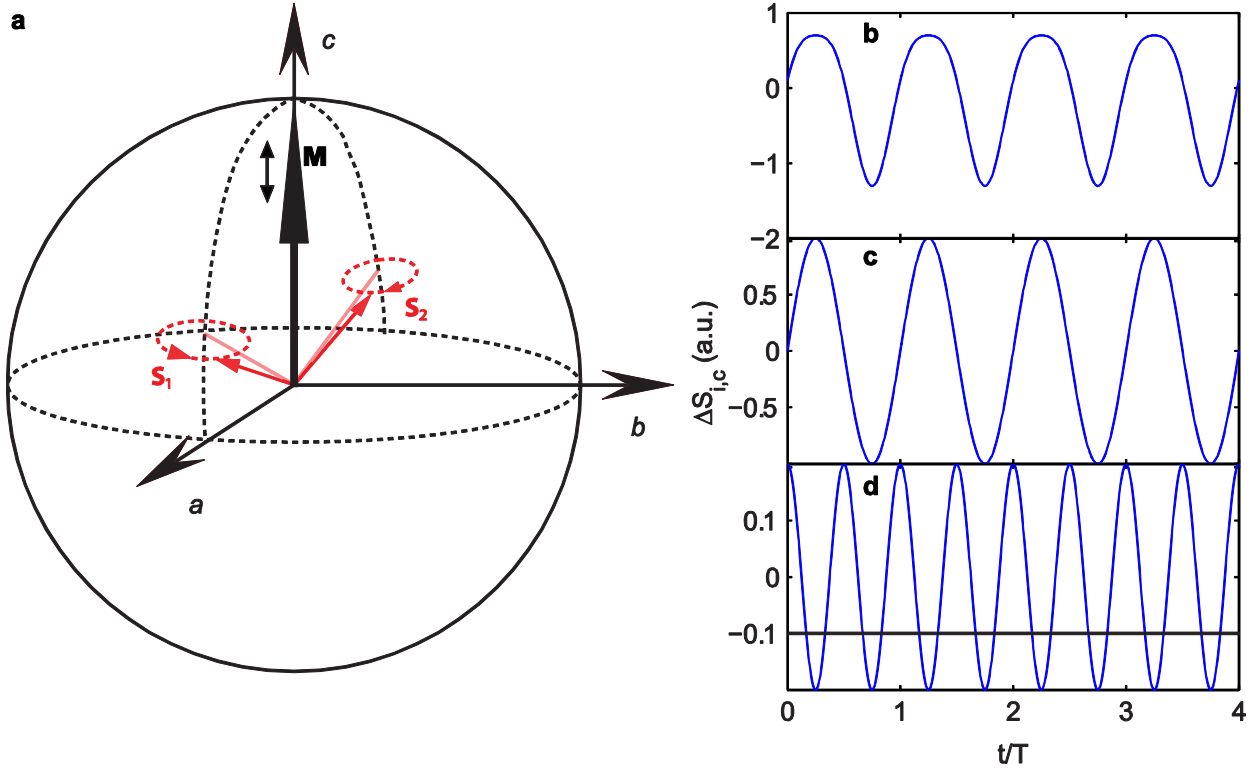

**Figure S8 | Temporal evolution of sublattice spins in AF magnon mode. a,** Exaggerated precession of sublattice spins $S_1$ and $S_2$ on the Poincaré sphere. The two red arrows denote $S_1$ and $S_2$ precessing around their static orientations indicated by the light red lines. The black arrow denotes the net magnetization **M**. The red dashed ellipses are the trajectories of $S_1$ and $S_2$ on the sphere surface. Note that under large angle precession, the excursion along $b$ axis is considerable, and the actual trajectories are distorted ellipses that are no longer planar. **b,** Temporal evolution of the deviations from the static projections of $S_1$ and $S_2$ along the crystal $c$ axis, $\Delta S_{i,c}$, as a function of time $t$ normalized to the precession period $T$. The projection along $c$ is periodic with a larger amplitude in the negative direction than in the positive direction (with respect to the projection of the static spin vectors) and results in a DC component in the negative $c$ direction. **c** and **d,** Fundamental (**c**), second harmonic (**d**, blue) and DC (**d**, black) components of $\Delta S_{i,c}$. This temporal evolution of $\Delta S_{i,c}$ gives rise to the second harmonic and rectification signals polarized along crystallographic $c$ axis radiated by $\mathbf{M}(t)$.

| Parameters | Values used[S5, S6] |
|---|---|
| Exchange constant $J$ | 4.24 (meV) |
| Antisymmetric exchange constant **D** | 0.066 (meV) |
| Crystalline anisotropy $K_a$ | 0.0069 (meV) |
| Crystalline anisotropy $K_c$ | 0.0046 (meV) |
| Total spin number $S$ | 5/2 |

**Table I | Details of the parameters used in numerical simulation.** The values of exchange constant $J$ and antisymmetric exchange constant **D** were adjusted slightly from literature values to match the magnon frequencies in the simulation to those observed in experiments.